\documentclass[aps,prl,twocolumn,superscriptaddress,showpacs]{revtex4-1}
\usepackage{hyperref}
\usepackage{amsmath}
\usepackage{graphicx}

\begin{document}

\title{Dynamical Screening Effect on Local Two-Particle Vertex Functions}
\author{Li Huang}
\affiliation{ Science and Technology on Surface Physics and Chemistry Laboratory, 
              P.O. Box 718-35, 
              Mianyang 621907, 
              Sichuan, 
              China }
\affiliation{ Department of Physics, 
              University of Fribourg, 
              Fribourg CH-1700, 
              Switzerland }

\author{Yilin Wang}
\affiliation{ Beijing National Laboratory for Condensed Matter Physics, 
              and Institute of Physics, 
              Chinese Academy of Sciences, 
              Beijing 100190, 
              China }

\date{\today}

\begin{abstract}
In principle, the electronic Coulomb interaction among the correlated
orbitals is frequency-dependent. Though it is generally believed that 
the dynamically screened interaction may play a crucial role in 
understanding the subtle electronic structures of strongly correlated 
materials, hitherto we know very little about it. In the Letter, we 
demonstrate that within the framework of single-site dynamical 
mean-field theory the local two-particle Green's functions $\chi$ and 
vertex functions $\Gamma$ are strongly modified by the dynamically 
screened interaction. Since both $\chi$ and $\Gamma$ represent the 
main ingredients to compute momentum-resolved response functions and 
to treat non-local spatial correlations by means of diagrammatic 
extensions of dynamical mean-field theory, it is urgent to reexamine 
previous results by taking the dynamical screening effect into 
account. The modifications should be very considerable.
\end{abstract}


\maketitle

Over the last years, tremendous progresses have been made in the studies
of strongly correlated materials by using the density functional theory
(DFT) combined with the dynamical mean-field theory (DMFT).\cite{antoine:13,kotliar:865}
The underlying physics in these materials can be generally captured by
effective Hamiltonian models which are derived in the DFT calculations. 
Subsequently, these models are solved numerically or analytically in the 
DMFT part to extract various one-particle and two-particle quantities, 
which are used to compare with the corresponding experimental results 
if available.\cite{va:2010}

In classical DFT + DMFT calculations, the electron-electron interactions 
among the correlated orbitals are usually modeled by Coulomb interaction 
matrix which is parameterized with Hubbard $U$. It is hardly surprising 
that the calculated results rely heavily on the style of Hubbard 
$U$.\cite{va:2010} Usually $U$ is assumed to be static. However, the 
charge density fluctuation in correlated orbitals will induce electric 
fields, which will be screened by the higher-lying or lower-lying states 
in the systems. The screening effect will finally result in a 
frequency-dependent renormalization for the Coulomb interaction. Thus, 
in principle the Coulomb interaction is dynamical, i.e, 
$U = U(\omega)$.\cite{arya:195104,arya:125106} Since the dynamical 
screening effect may exist in strongly correlated materials widely, how 
to treat it is one of the major challenges in studying the intriguing
properties of these materials. Due to many technical difficulties, it has 
been completely neglected or empirically taken into account by adjusting 
the static value $U_{0} = U(\omega = 0)$\cite{va:2010} or reducing the 
effective bandwidth of correlated orbitals\cite{casula:2012} in most calculations.
Very recently, numerous efforts have been made to explore the dynamical 
screening effect in strongly correlated materials,\cite{werner:146401,
werner:146404,assaad:035116,casula:2011} but very little is known on the 
consequences of frequency-dependent Coulomb interaction. So far the only 
thing we can confirm is that the dynamical screening effect may play an 
essential role in understanding the fascinating properties of strongly 
correlated materials,\cite{ta:226401,ta:12102712} such as the plasmon 
satellites and spectral weight transfer in the photoemission spectra of 
SrVO$_{3}$\cite{jm:67001,lh:67003} and hole-doped BaFe$_{2}$As$_{2}$,\cite{werner:331337} 
which are less emphasized before.

The local two-particle vertex functions represent a crucial ingredient 
for the calculations of (momentum-dependent) dynamical susceptibilities, 
such as spin density wave, optical conductivity, thermopower, 
and thermal conductivity, etc.\cite{boe:115128,park:137007,nl:106403,
boe:075145,jk:085102,hh:205106,gr:125114} Remarkably, it is argued that
the local vertex functions can be used to calculate the electric polarization
in correlated insulators\cite{rn:2013} and to identify the fingerprints 
of Mott-Hubbard metal-insulator transition.\cite{ts:2013} Beyond that, 
the local two-particle vertex functions are also the prerequisites for  
the diagrammatic extensions of the DMFT, such as the dynamical vertex 
approximation (D$\Gamma$A)\cite{slezak:435604,toschi:045118} and dual 
fermion (DF)\cite{rubtsov:033101} approach, which aiming at the inclusion 
of non-local spatial correlations. Therefore, to obtain reliable and accurate
local two-particle vertex functions is one of the major tasks in DMFT 
calculations all the time. However, though many numerical tricks and 
approximations were developed in recent years,\cite{boe:075145,hh:205106,
jk:085102} it is still higher difficulty and heavier workload of performing 
calculations at the two-particle level than the one-particle level. As
a result it is not astonished that the analysis of two-particle quantities 
is usually restricted to the easiest cases of optical conductivity, for 
which one can neglect the vertex correction and retain the bubble term 
only, and then the calculation can be simplified to one-particle level 
eventually.\cite{antoine:13,kotliar:865,va:2010}

In this Letter, we calculate and analyze the local two-particle Green's
functions $\chi$ and vertex functions $\Gamma$ for single-band Hubbard 
model and strongly correlated metal SrVO$_{3}$ by using the DMFT and DFT 
+ DMFT approaches\cite{antoine:13,kotliar:865,va:2010} respectively. The 
dynamical screening effects for both the Hubbard model and realistic materials 
are considered on the same footing in our calculations. We found that the 
calculated results for both $\chi$ and $\Gamma$ are significantly affected 
by the dynamically screened $U$. Therefore, it is suggested that the previous 
results and conclusions, concerning with the momentum-resolved response 
functions and diagrammatic extensions of DMFT, etc., should be reexamined 
and reconsidered carefully, provided that the dynamical screening effect 
is sizable and can not be ignored completely.

\begin{figure}
\centering
\includegraphics[scale=0.64]{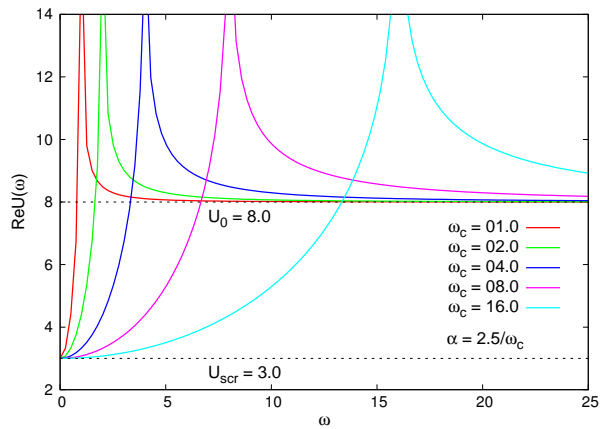}
\caption{(Color online) The typical Ohmic models used in the present 
calculations for frequency-dependent Coulomb interaction $\text{Re} 
U(\omega)$. The $U_{0}$ and $U_{\text{scr}}$ are fixed to 8.0 eV and 3.0
eV, respectively. \label{fig:model}}
\end{figure}

We now illustrate the analytical Ohmic model\cite{werner:146401} which 
is often used to describe the dynamically screened behaviour for Hubbard 
$U$ :
\begin{equation}
\text{Re} U(\omega) = U_{0}
+ \alpha \omega \ln \left|\frac{\omega_{c} + \omega}{\omega_{c} - \omega}\right| 
- 2\alpha\omega_{c},
\end{equation}
where $\alpha$ and $\omega_{c}$ are the screening parameters, $U_{0}$ 
is the static interaction, and the screened interaction can be defined 
as $U_{\text{scr}} = U_{0} - 2\alpha \omega_{c}$. Note that a characteristic 
feature of $U(\omega)$ of paramagnetic Ni can be captured by this Ohmic 
model.\cite{arya:195104,arya:125106} Several typical Ohmic models used 
in the present calculations are shown in Fig.\ref{fig:model}. In order 
to include the dynamical screening effect into the DMFT and DFT + DMFT 
frameworks,\cite{werner:146404,assaad:035116,casula:2011} the hardest 
obstacle has been the lack of a reliable and efficient quantum impurity 
solver for the general impurity model with a frequency-dependent Coulomb 
interaction $U(\omega)$. Fortunately, it seems that this obstacle has 
been overcome by the recent development of hybridization expansion 
continuous-time quantum Monte Carlo impurity solver (dubbed CT-HYB)
proposed by Werner \emph{et al.},\cite{werner:146401,werner:146404} 
where a multi-plasmon Lang-Firsov transformation is treated exactly 
in the context of a hybridization expansion algorithm for the general 
impurity model.

According to the literatures,\cite{gr:125114,hh:205106} the general
two-particle Green's function $\chi$ can be defined as follows:
\begin{equation}
\chi_{abcd}(\tau_a,\tau_b,\tau_c,\tau_d) = 
\langle T_{\tau} c_a(\tau_a) c^{\dagger}_b(\tau_b) c_c(\tau_c) c^{\dagger}_d(\tau_d) \rangle.
\end{equation}
Generally these functions are evaluated in frequency space, as a function
of two fermionic frequencies ($\nu$ and $\nu'$) and one bosonic frequency 
($\omega$). Then the local two-particle vertex functions 
$\Gamma^{\nu\nu'\omega}_{\sigma\sigma'}$ can be derived directly by 
$\chi^{\nu\nu'\omega}_{\sigma\sigma'}$ through the following equation:
\begin{widetext}
\begin{equation}
\chi^{\nu\nu'\omega}_{\sigma\sigma'} = 
-\beta G_{\sigma}(\nu) G_{\sigma}(\nu + \omega) \delta_{\nu\nu'}\delta_{\sigma\sigma'}
-G_{\sigma}(\nu)G_{\sigma}(\nu + \omega)
 \Gamma^{\nu\nu'\omega}_{\sigma\sigma'}
 G_{\sigma'}(\nu')G_{\sigma'}(\nu' + \omega),
\end{equation}
\end{widetext}
where $\beta$ is inverse temperature.
We calculate the $\chi^{\nu\nu'\omega}_{\sigma\sigma'}$ and 
$\Gamma^{\nu\nu'\omega}_{\sigma\sigma'}$ with single-site DMFT and 
DFT + DMFT methods by using the powerful CT-HYB quantum impurity 
solver.\cite{werner:076405,gull:349} An improved estimator for 
$\chi^{\nu\nu'\omega}_{\sigma\sigma'}$ suggested by Hafermann 
\emph{et al}.\cite{boe:075145,hh:205106} is adopted to suppress the 
numerical noises and yield accurate data.

For the sake of simplicity, above all we consider a half-filled one-band 
Hubbard model on the Bethe lattice with bandwidth $4t$ at $\beta = 50.0$. 
The static and screened Hubbard $U$ are 8.0 eV and 3.0 eV, respectively. 
Then we tune the $\alpha$ and $\omega_{c}$ parameters for the Ohmic model 
to monitor the variational trends for selected one-particle quantities and 
local two-particle vertex functions.
 
\begin{figure}
\centering
\includegraphics[scale=0.64]{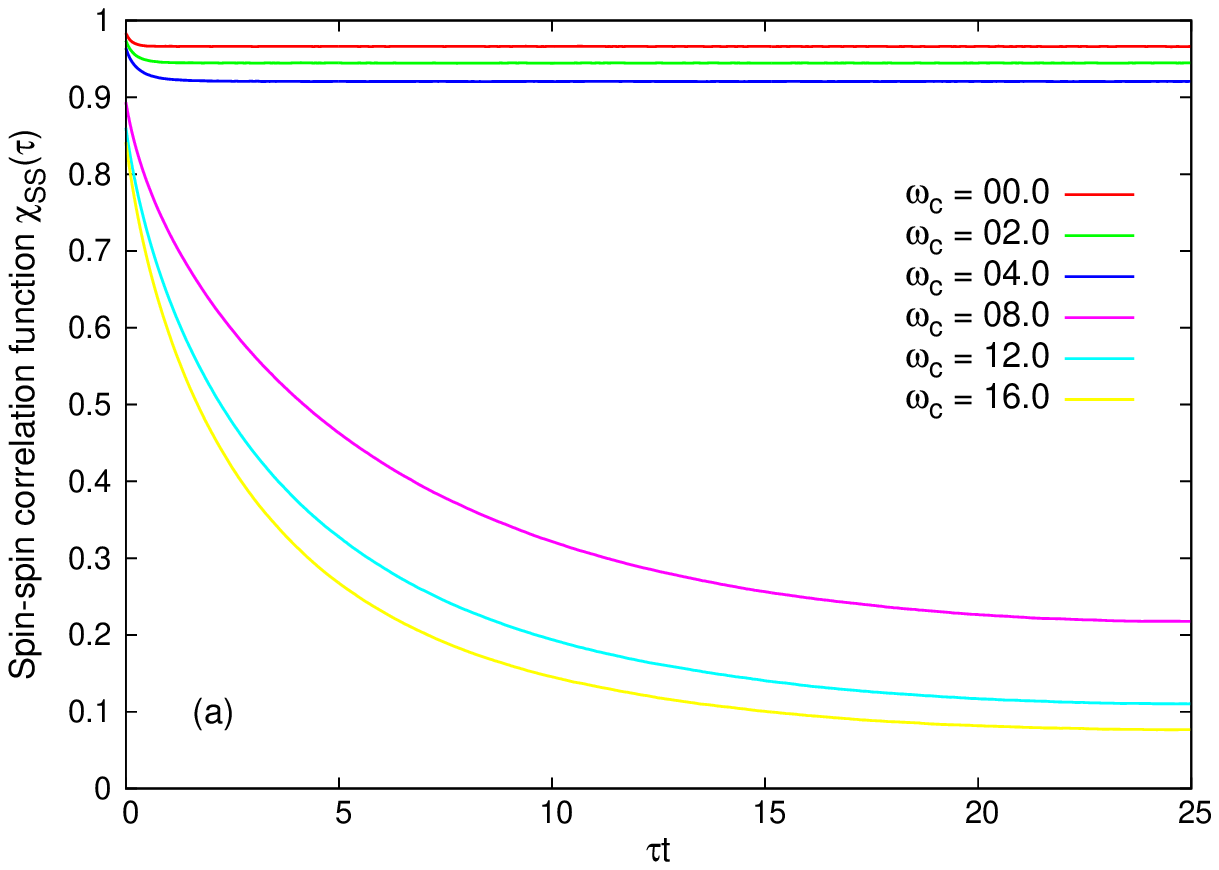}
\includegraphics[scale=0.64]{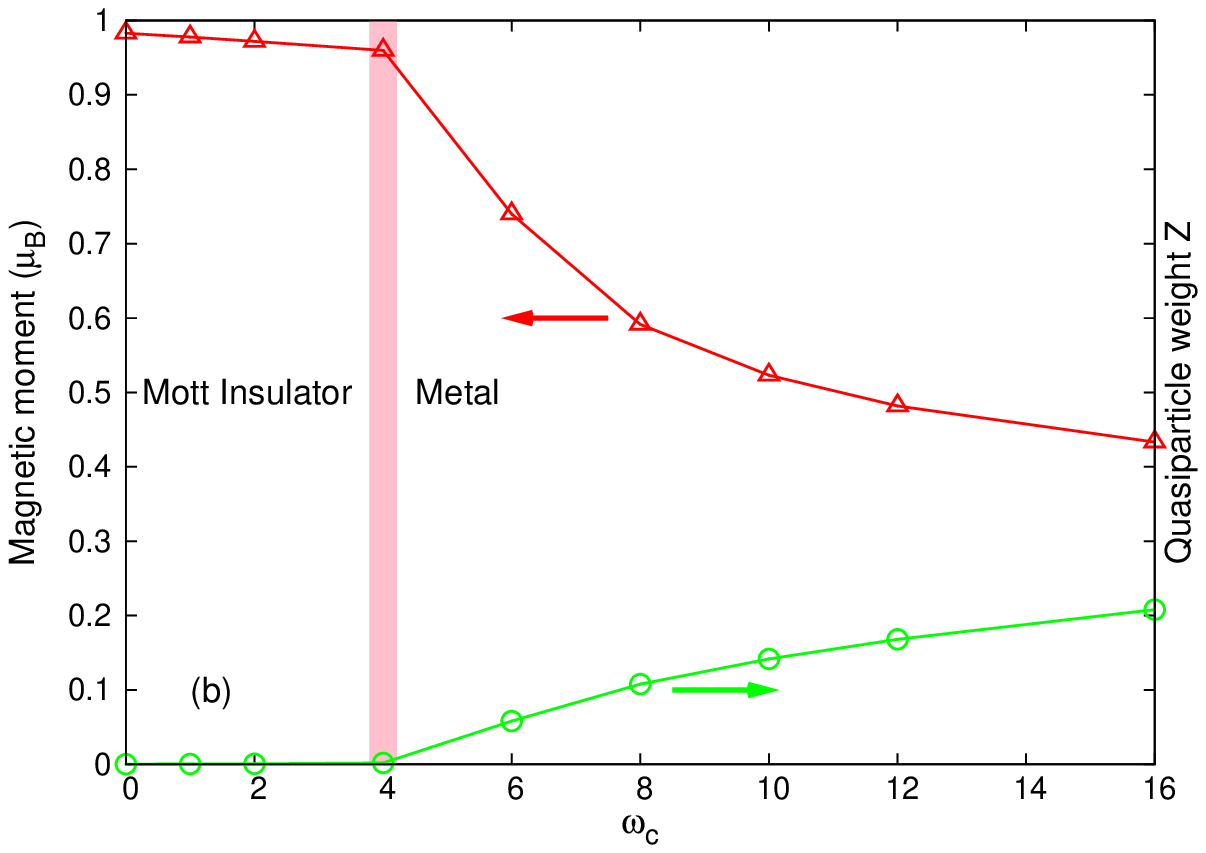}
\caption{(Color online) Magnetic moment collapse and metal-insulator transition 
in half-filled one-band Hubbard model induced by the dynamical screening effect.
(top) Spin-spin correlation function $\chi_{\text{SS}}(\tau) = \langle S_{z}(0) 
S_{z}(\tau) \rangle$. (bottom) Quasiparticle weight $Z$ and local spin magnetic 
moment $\sqrt{T \chi_{\text{SS}}}$ as a function of $\omega_{c}$. The Mott 
metal-insulator transition zone is highlighted by a pink vertical bar.\label{fig:schi}}
\end{figure}

(a) \emph{Dynamical screening effect on spin-spin correlation function 
and local magnetic moment in one-band Hubbard model.} 
Previous studies show that the dynamically screened interaction can 
shift the metal-insulator phase boundary,\cite{werner:146401} change 
the spectral function near the Mott-Hubbard gap edge, and renormalize 
the quasiparticle band structure.\cite{lh:67003,jm:67001} Inspired by 
these works, we study the spin-spin correlation function 
$\chi_{\text{SS}}(\tau) = \langle S_{z}(0) S_{z}(\tau) \rangle$ in 
the Hubbard model with frequency-dependent $U$ to gain more insight 
into the dynamical screening effect. The obtained $\chi_{\text{SS}}(\tau)$, 
effective local magnetic moment $M_{e} = \sqrt{T \chi_{\text{SS}}}$, 
and quasiparticle weight $Z$ are shown in Fig.\ref{fig:schi}. As is seen 
in this figure, the $\chi_{\text{SS}}(\tau)$, $M_{e}$, and $Z$ are
strongly modulated by the screening frequency parameter $\omega_{c}$
of the Ohmic model. Obviously, there exists a critical $\omega_c$ 
separating the metal and Mott insulator phases. We just focus on the 
following two cases:
(i) The screening frequency $\omega_{c}$ is small ($\omega_{c} \leq 4.0$).
The $\chi_{\text{SS}}(\tau)$ approaches a non-zero constant at large times 
which characterizes a frozen moment phase. $M_{e}$ only decreases slightly 
from its initial value ($0.98\ \mu_{B}$ at $\omega_{c} = 0.0$) under the 
increment of $\omega_{c}$. And the quasiparticle weight $Z$ approximates 
zero, which means the system should exhibit completely insulating properties.
(ii) The screening frequency $\omega_{c}$ is large ($\omega_{c} > 4.0$).
The behavior of $\chi_{\text{SS}}(\tau)$ manifests that the system still 
retains the frozen moment phase. Whereas, $M_{e}$ decreases quickly and 
monotonously from $0.96\ \mu_{B}$ to $0.43\ \mu_{B}$, i.e, the effective 
local magnetic moment appears apparent collapse. On the other hand, the 
quasiparticle weight $Z$ becomes more and more considerable when $\omega_{c}$ 
increases, which means that the system goes into a fully metallic state.
Further analysis on the imaginary part of the calculated low-frequency 
Matsubara self-energy function $\text{Im} \Sigma(i\omega)$ indicates that 
it exhibits noticeably fractional power-law behavior with respect to 
$i\omega$, which is the signature of the 
non-Fermi-liquid metallic phase. In Fig.\ref{fig:schi}, we observe a 
sudden drop for $M_{e}$ and rise for $Z$ around $\omega_{c} = 4.0$, which 
corresponds to a typical first-order transition.


\begin{figure}
\centering
\includegraphics[scale=0.64]{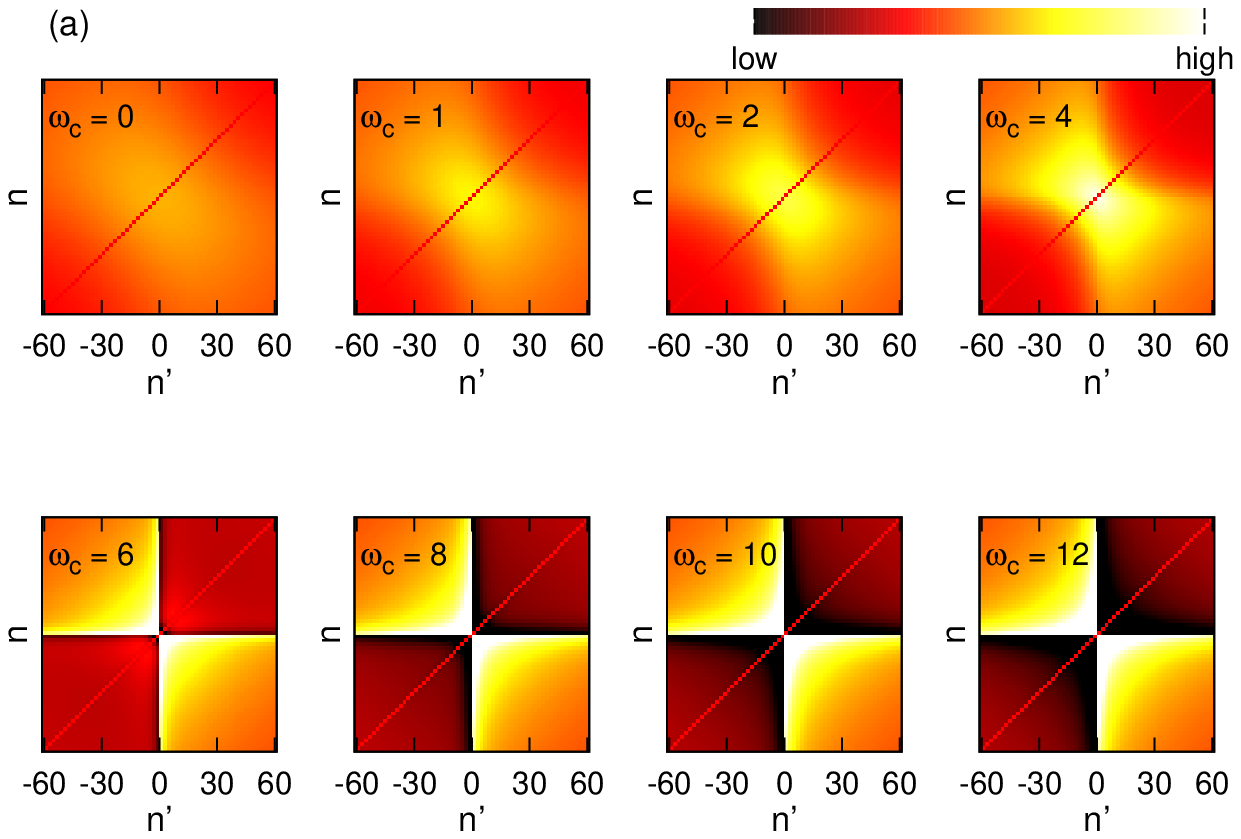}
\includegraphics[scale=0.64]{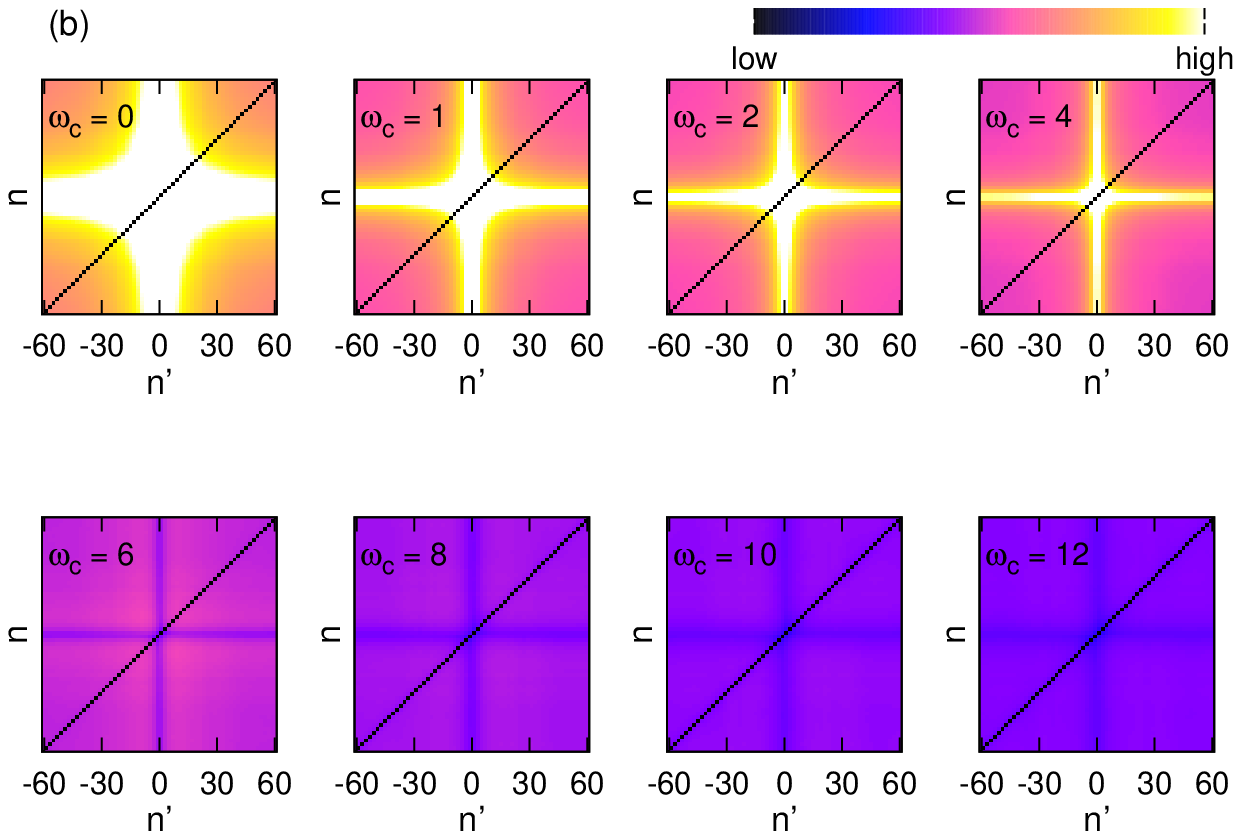}
\caption{(Color online) Vertex functions vs. the two fermionic frequencies
$\nu = (2n + 1) \pi /\beta$ and $\nu' = (2n'+1)\pi/\beta$ as a function of 
the screening frequency $\omega_{c}$ for half-filled single-band Hubbard model.
(top) Real part of spin-up-up component for the local two-particle Green's 
functions: 
$\text{Re} \chi^{\nu\nu'\omega}_{\uparrow\uparrow}$.
(bottom) Real part of spin-up-up component for the local two-particle vertex 
functions:
$\text{Re} \Gamma^{\nu\nu'\omega}_{\uparrow\uparrow}$.
Note that the bosonic frequency $\omega$ is fixed to be zero. Instead of 
the absolute values of the Matsubara frequencies just the corresponding
indexes are given so as to improve the readability of these figures.
\label{fig:v_t}}
\end{figure}

(b) \emph{Dynamical screening effect on local two-particle vertex functions 
in one-band Hubbard model.}
We then concentrate our attentions to the local two-particle Green's functions
$\chi^{\nu\nu'\omega}_{\sigma\sigma'}$ and full vertex functions 
$\Gamma^{\nu\nu'\omega}_{\sigma\sigma'}$. In Fig.\ref{fig:v_t}, only 
the contour maps for their spin-up-up components (the bosonic frequency 
$\omega = 0$) are displayed. As is clear in this figure, if the dynamical 
screening effect is included in the DMFT calculations, not only the 
one-particle quantities but also the two-particle quantities are modified 
visibly, which are not taken seriously before.
Let's look at the $\text{Re} \chi^{\nu\nu'\omega}_{\uparrow\uparrow}$ part 
(see top panel in Fig.\ref{fig:v_t}) at first. The intensity plots of it 
can be divided into two kinds. 
(i) For $\omega_{c} \leq 4.0$, one observes a gradual enhancement at the 
low-frequency zone (central area). But neither the singularity nor obvious 
geometric structures is detected.
(ii) For $\omega_{c} > 4.0$, a numerical divergence emerges at $\nu \rightarrow 
0$ and $\nu' \rightarrow 0$ zone, where the intense white-black color coding 
indicates alternating signs in the ($\nu$,$\nu'$) space. This low-frequency 
divergence becomes more and more prominent when $\omega_{c}$ increases 
steadily. In addition, one can distinguish two ``cone''-like structures 
easily which are totally absent when $\omega_{c} \leq 4.0$.
Next we pay attention to the $\text{Re} \Gamma^{\nu\nu'\omega}_{\uparrow\uparrow}$
part (see bottom panel in Fig.\ref{fig:v_t}). The most striking features 
in the contour plots is the cross structure. Interestingly, the vertices 
can be split into two types based on the intensity and sign of the cross
structures as well.
(i) For $\omega_{c} \leq 4.0$, the cross structure is notable and positive.
Upon increasing the screening frequency $\omega_{c}$, the intensity
weakens and the extent shrinks. 
(ii) For $\omega_{c} > 4.0$, the cross structure exists too, but its
sign is now inverted as indicated by the colors. In other words, the
cross structures are concave, contrary to those in the $\omega_{c} \leq 
4.0$ case. Furthermore, the vertex tends to be featureless with the 
increment of $\omega_{c}$.
Thus, it is suggested that the frozen moment phase with large local 
magnetic moment will strengthen the two-particle vertex functions $\Gamma$, 
which is in accord with the previous results obtained by Hafermann 
\emph{et al}.\cite{hh:205106} They found that the vertex is essentially 
featureless on the Fermi liquid side, and evident in the frozen moment 
phase. To make a long story short, dramatic changes will occur in both the 
intensity and structure of $\text{Re}\chi^{\nu\nu'\omega}_{\uparrow\uparrow}$ 
and $\text{Re}\Gamma^{\nu\nu'\omega}_{\uparrow\uparrow}$, accompanying with 
the first-order metal-insulator transition and collapse of local magnetic 
moment, when the screening frequency $\omega_{c}$ approaches the critical point. We also 
make detailed analysis on the other components or channels for $\chi$ and 
$\Gamma$, and the other reducible and irreducible vertex functions.\cite{gr:125114} 
They exhibit analogous features when the dynamical screening effect is 
activated.


\begin{figure}
\centering
\includegraphics[scale=0.64]{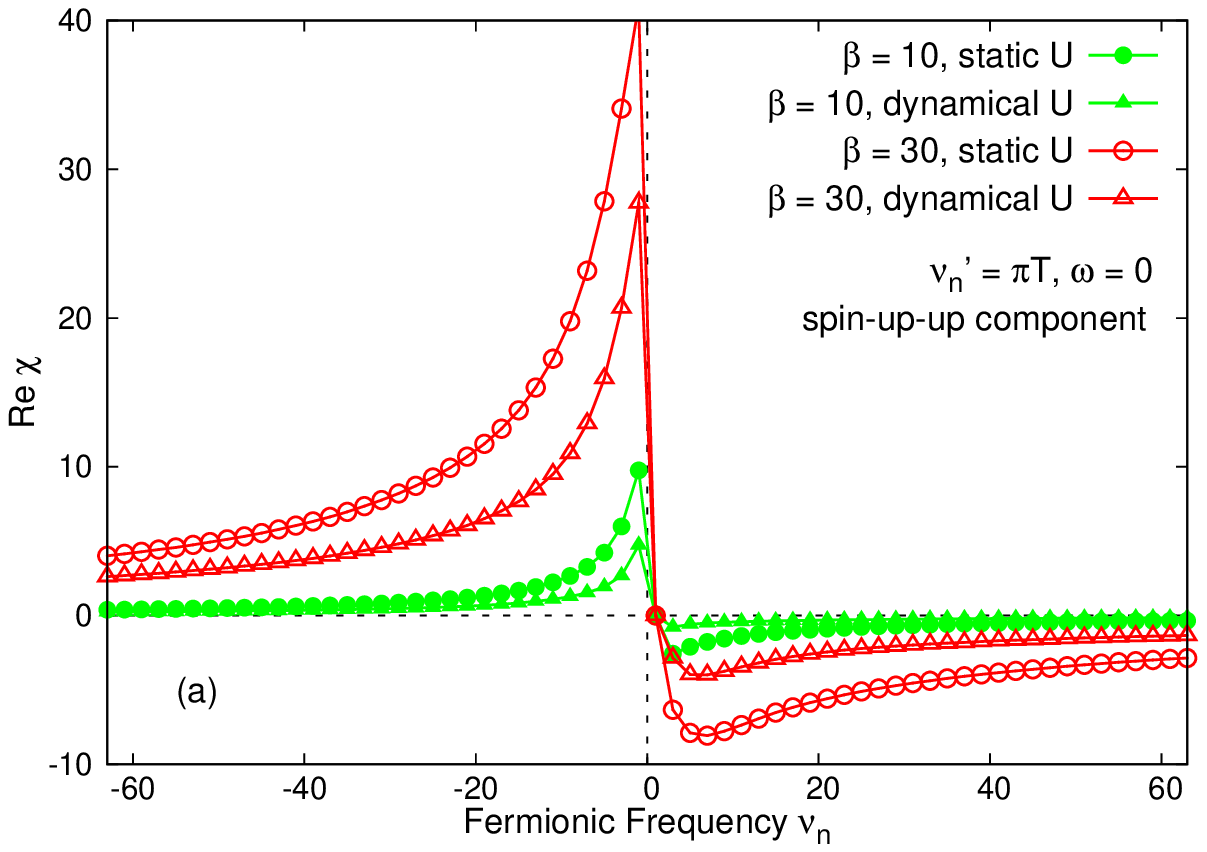}
\includegraphics[scale=0.64]{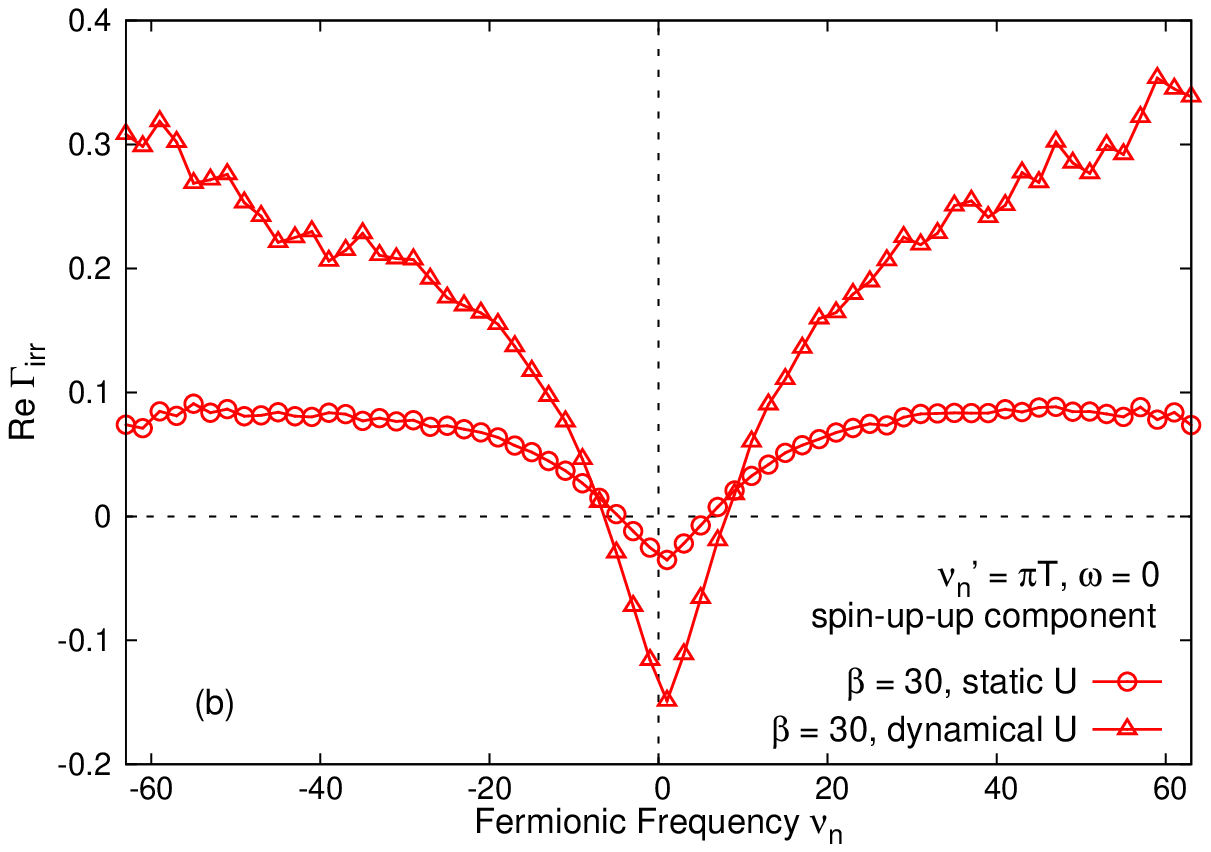}
\caption{(Color online) Real part of the spin-up-up component of vertex 
functions for strongly correlated metal SrVO$_{3}$ obtained in the 
DFT + DMFT calculations. (top) Local two-particle Green's functions 
$\text{Re} \chi^{\nu\nu'\omega}_{\sigma\sigma'}$. (bottom) Local irreducible 
vertex functions $\text{Re} \Gamma_{\text{irr}}$. The second fermionic 
frequency $\nu'$ is fixed to be $\pi T$, and the bosonic frequency $\omega$ 
is set to be zero. The static $U$ is 4.0\ eV, and the data for frequency-dependent 
$U$ are taken from reference \cite{arya:125106} directly. The DFT + DMFT 
calculations are done for $\beta = 10$ and $\beta = 30$, respectively.
\label{fig:v_t_svo}}
\end{figure}

(c) \emph{Dynamical screening effect on local two-particle vertex 
functions in SrVO$_{3}$.}
So far all the discussions are limited to the simplest single-band Hubbard 
model, and the frequency-dependent interaction $U$ is modeled by analytical 
Ohmic model.\cite{werner:146401} In order to confirm the above conclusions, 
we conduct further DFT + DMFT calculations for strongly correlated metal 
SrVO$_{3}$ to examine the consequences of dynamical screening effect on the 
two-particle Green's functions and vertex functions. SrVO$_{3}$ is an ideal 
system to benchmark the new theories and tools describing the strongly 
correlated materials. It has been the subject of numerous experimental and 
theoretical studies.\cite{va:2010} In previous works, several DFT + DMFT 
calculations were devoted to study the one-particle quantities of SrVO$_{3}$ 
under the dynamical screening effect.\cite{casula:2011,jm:67001,lh:67003,casula:2012}
It is shown to renormalise the spectral weight near the Fermi level, to 
increase the effective mass, and to suppress the $t_{2g}$ quasiparticle 
bandwidth distinctly. Nevertheless, to our knowledge, the two-particle 
quantities of SrVO$_{3}$ have not been studied before neither experimentally 
nor theoretically. So our results can be regarded as a valuable prediction.

The frequency-dependent interaction for realistic materials can be evaluated 
at the random phase approximation (RPA) level as usual. As for SrVO$_{3}$, 
the $U(\omega)$ data are taken from reference \cite{arya:125106} directly. 
The details for our DFT + DMFT calculations can be found in reference \cite{lh:67003}.
The main calculated results, real part of spin-up-up component of local 
two-particle Green's function $\text{Re} \chi^{\nu\nu'\omega}_{\sigma\sigma'}$, 
are shown in the top panel of Fig.\ref{fig:v_t_svo}. Comparing the calculated 
results with and without dynamical interaction, the difference is rather 
significant, especially for the $\nu \rightarrow 0$ region. Moreover, the difference 
is amplified at lower temperature. When the dynamical screening effect is 
included, the absolute values of $\text{Re} \chi^{\nu\nu'\omega}_{\sigma\sigma'}$ 
are strongly reduced. Similar or opposite trends are found for the other 
two-particle vertex functions, such as the local irreducible vertex functions 
$\Gamma_{\text{irr}}$ (see the bottom panel of Fig.\ref{fig:v_t_svo}).
The vertices $\Gamma_{\text{irr}}$ can be evaluated straightforward through 
the famous Bethe-Salpeter equation: $ \Gamma_{\text{irr}} = \beta (\chi^{-1}_{0} 
- \chi^{-1})$, where the polarization bubble $\chi_{0}$ is computed 
from the fully interacting one-particle Green's function.\cite{gr:125114}

Finally, it should be stressed that in many strongly correlated materials, 
the dynamical screening effect is conspicuous.\cite{arya:195104,arya:125106} 
Werner \emph{et al}.\cite{werner:331337} have demonstrated that the fine 
electronic structures of hole-doped BaFe$_{2}$As$_{2}$ can be fully understood 
only if the dynamical screening effect is taken into account for the first 
time. Recently, Park \emph{et al}.\cite{park:137007} have studied the magnetic 
excitation spectra in BaFe$_{2}$As$_{2}$ through a two-particle approach 
within standard DFT + DMFT framework. Their calculated results, including 
the local irreducible vertex $\Gamma_{\text{irr}}$, dynamic magnetic 
susceptibility $\chi(\textbf{q},i\omega)$, and dynamical structure factor 
$S(\textbf{q},\omega)$, roughly reproduced all the experimentally observed 
features in inelastic neutron scattering.\cite{liu:2012} Since in their 
calculations just the static interaction was used, we believe that incorporating 
the dynamical screening effect into the DFT + DMFT calculations will contribute 
to improving their results.

In summary, we study several typical one-particle and two-particle quantities 
in both the half-filled single-band Hubbard model and transition metal oxide 
SrVO$_{3}$ by applying the DMFT and DFT + DMFT methods with frequency-dependent 
interaction $U$. In the case of the Hubbard model, the dynamical screening effect is shown to 
drive first-order metal-insulator transition and magnetic moment collapse by 
tunning the screening parameter $\omega_{c}$ of the Ohmic model. The intensity 
and structure for both the local two-particle Green's functions 
$\chi^{\nu\nu'\omega}_{\sigma\sigma'}$ and full vertex functions 
$\Gamma^{\nu\nu'\omega}_{\sigma\sigma'}$ display dramatic changes when the 
critical $\omega_{c}$ is attained. Concerning the strongly correlated 
metal SrVO$_{3}$, the calculated two-particle vertex functions are also 
suppressed or enhanced by the dynamical screening explicitly. Enormous changes 
for the \textbf{q}-dependent response functions are expected. Anyway, the 
two-particle quantities in either Hubbard model or realistic materials are 
intensively affected by the dynamical screening effect. Hence, in future 
studies with regards to moment-resolved response functions and diagrammatic 
extensions of DMFT, careful considerations for the dynamical screening  
effect should be essential.

\begin{acknowledgments}
We acknowledge financial support from the National Science Foundation
of China and by the 973 program of China (No.2013CBP21700 and No.2011CBA00108). 
All of the DMFT and DFT + DMFT calculations have been performed on the 
TIANHE-1A at National Supercomputing Center of Tianjin (NSSC-TJ).
\end{acknowledgments}

\bibliography{vertex}

\end{document}